\documentclass[titlepage,12pt]{utarticle}
\usepackage{amssymb, amsmath, amsopn, amsthm,amscd,amsfonts}
\usepackage{color}
\usepackage{latexsym}
\usepackage{ifthen}
\usepackage{epsfig}
\usepackage{graphicx}
\usepackage{graphics}
\usepackage{bm,longtable,enumerate}
\numberwithin{equation}{section}
\newcommand{\Tr}{{\rm Tr}}
\newcommand{\tr}{{\rm tr}}

\newcommand{\ol}{\overline}
\newcommand{\del}{\partial}

\newcommand{\nn}{\nonumber}
\newcommand{\half}{\frac{1}{2}}

\newcommand{\bR}{\mathbb{R}}
\newcommand{\eqn}{\begin{eqnarray}}
\newcommand{\eqnx}{\end{eqnarray}}

\newcommand{\bZ}{\mathbb{Z}}
\def\beq{\begin{equation}}
\def\eeq{\end{equation}}
\def\beqa{\begin{eqnarray}}
\def\eeqa{\end{eqnarray}}
\def\ss{\scriptscriptstyle}

\def\tr{\mathop{\rm tr}\nolimits}
\def\Tr{\mathop{\rm Tr}\nolimits}

\newcommand{\bH}{\mathbb{H}}
\newcommand{\ba}{{\boldsymbol{a}}}

\newcommand{\bk}{{\vec{k}}}
\newcommand{\bp}{{\vec{p}\,}}
\newcommand{\bX}{{\vec{X}}}

\newcommand{\psl}{\not\! p}

\def\matt[#1,#2,#3,#4]{\left(%
\begin{array}{cc} #1 & #2 \\ #3 & #4 \end{array} \right)}

\def\v2#1{\vv2[#1]}
\def\vv2[#1,#2]{\left(%
{#1 \atop #2}\right)}

\def\ol{\overline}

\def\nn{\nonumber}

\def\eps{\epsilon}

\begin{document}
\preprint{DIAS-STP-11-09\\}
\title{{\bf Relativistic baryons in the Skyrme model revisited}}
\author{ {\sc Henrique Boschi-Filho $^{\rm a}$ , Nelson R.~F.~Braga} 
 \address{
 Instituto de F{\'i}sica, \\
 Universidade Federal do Rio de Janeiro,\\
         Caixa Postal 68528,\\
         21941-972 Rio de Janeiro, RJ, Brasil.\\[0.2cm]
         Email: {\tt boschi@if.ufrj.br},\\
         ~~~~~~~~~{\tt braga@if.ufrj.br},\\
         ~~~~~~~~~{\tt mtorres@if.ufrj.br}
          } ,
 {\sc Matthias Ihl}
 \address{
        School of Theoretical Physics, \\
        Dublin Institute for Advanced Studies, \\
        10 Burlington Rd, \\
        Dublin 4, Ireland. \\[0.3cm]
        Email: {\tt msihl@stp.dias.ie}
         } ,
 \\ {\sc Marcus A.~C.~Torres} $^{\rm a}$  
}

\Abstract{We investigate and revisit certain aspects of relativistic baryons in the Skyrme model. Starting from static baryons in the helicity eigenstates, we generalize their wavefunctions to the non-static and relativistic regime. 
A new representation for gamma matrices in the SU(2) collective space is constructed and the corresponding Dirac equation is obtained. 
Furthermore, we extend the analysis of Adkins {\it et al.} concerning vector current vacuum expectation values for baryon states to arbitrary momenta and calculate baryon elastic form factors 
in a general framework exhibiting vector meson dominance.}

\maketitle
\newpage

\section{Introduction}

The Skyrme ($SU(2)\times SU(2)$ chiral) model \cite{Skyrme:1961vq} has attracted renewed interest as a simplified description of baryons in the large $N_c$ limit of QCD.
The static properties of baryons in the Skyrme model (skyrmions) have been studied a long time ago \cite{Adkins:1983ya, Adkins:1983hy}.
Some properties of slowly moving skyrmions have been studied in various publications, e.g., \cite{Braaten:1986iw, Holzwarth:2005re}.\\
In \cite{Sakai:2004cn}, 
Sakai and Sugimoto derive the Skyrme model as the $3+1$ dimensional pion effective theory descending from the dynamics of the gauge fields living on flavor $D8$-branes that probe
the $D4$-brane geometry generated by a stack of color branes. 
The basic idea is to treat baryons in holographic theories as 
solitons in an effective theory of mesons which at low energies reduces to a non-linear sigma model with broken chiral symmetry. Further restriction to 
two flavors only, i.e. $SU(2)$ flavor symmetry, results in a vanishing Wess-Zumino term. This yields precisely the above-mentioned Skyrme model. \\
The purpose of this letter is to calculate non-static baryon wave functions in their helicity eigenstates and generalize the results of \cite{Adkins:1983ya} to the relativistic case.
This will be presented in a form compatible with the Dirac spinor representation that eventually will allow us to consider relativistic 
baryons by boosting in any direction. Here, we are disregarding relativistic corrections to the non-relativistic Lagrangian obtained by quantizing slowly moving baryons
via the moduli space approximation method. Moreover, we are ignoring other possible effects and corrections such as deformations of spinning Skyrmions \cite{Hata:2010vj, Hata:2010zy}. Related work discussing corrections to the Lagrangian of the collective states can also be found in \cite{Ji:1991ff}. 
The approach discussed in this note provides a relativistic description of the baryon wavefunctions that allows us to correctly decompose the matrix elements of baryonic currents and properly define the associated form factors.

\section{From Skyrmions to Instantons (and back)}

In this section we briefly review the Skyrme theory and its link to instantons and holography. We follow closely the previous works \cite{Atiyah:1989dq,Sutcliffe:2010et}.
Skyrmions are soliton solutions of a nonlinear effective field theory of pions. The model is nonrenormalizable and holographic models \cite{Sakai:2004cn, Hata:2007mb,Hashimoto:2008zw, BallonBayona:2009ar, Bayona:2010bg, Ihl:2010zg} provide a UV completion. 
The action of the Skyrme model is given by
\begin{eqnarray}
S=\int d^4 x\left(\frac{f_\pi^2}{4}\tr\left(
 U^{-1}\del_\mu U\right)^2+\frac{1}{32 e^2}
\tr\left[U^{-1}\del_\mu U,U^{-1}\del_\nu U\right]^2
\right) \, 
\end{eqnarray}
where the pion fields $\pi(x^\mu)$ are encoded in the SU(2) valued Skyrme field,
\beq
U(x^\mu)= e^{2i\pi(x^\mu)/f_\pi}.
\eeq 
The Skyrmion is a static solution that goes to a constant value at infinity which compactifies the space to $S^3$, and a topological charge $n_B \in \pi_3(SU(2))= {\mathbb Z}$ is identified with 
the baryon number,
\beq
n_B = -\frac{1}{24 \pi^2} \int \epsilon_{ijk} {\mathrm Tr}\left(R_i R_j R_k \right) d^3 x,
\eeq
where $R_i = \partial_i U \; U^{-1}$.
It is worth noting that the Skyrmion field exceeds the BPS bound,
\beq
E_{\mathrm{Skyrmion}}\geq 12 \pi^2 |n_B|,
\eeq
by 23$\%$ for $n_B=1$, but gets closer to the bound for Skyrmion crystal field configurations \cite{Klebanov:1985qi,Kugler:1988mu} for large $n_B$. \\
In a seminal publication, Atiyah and Manton \cite{Atiyah:1989dq} showed that (static) skyrmions can be approximated remarkably well by generating fields $U_{\mathrm{inst.}}$ analytically from the
holonomy of a SU(2) Yang-Mills instanton $A_I$ on $\mathbb{R}^4$, i.e.,
\beq
U_{\mathrm{inst.}}(\vec{x}) = \pm {\mathcal P} \mathrm{exp} \int_{-\infty}^{+\infty} A_4  (\vec{x}, x_4) dx_4,
\eeq
where $\mathcal P$ denotes path ordering, $\vec{x} \in \mathbb{R}^3$, and $x_4$ denotes Euclidean time. In holographic models \cite{Hata:2007mb,Hashimoto:2008zw,Pomarol:2007kr}, $x_4$ denotes the holographic (radial) direction. Then one can show that the instanton number precisely agrees with the 
baryon number $n_B$ defined above and that, for a suitable choice of scale, the skyrmions are approximated very well (the energies usually lie within a percent of the corresponding 
numerical Skyrme solution).\\  
Another significant extention of the  Skyrme model is the Hidden Local Symmetry approach \cite{Bando:1987br}. The Skyrme model is a theory of pions that lacks the presence of other mesons as would be expected in the large $N_c$ limit of QCD. In the Hidden Local Symmetry model, the diagonal $SU(2)$ subgroup of the chiral symmetry $SU(2)_L\times SU(2)_R$ is gauged and gives rise to vector mesons. \\ 
In holographic models, the effective action on the probe brane typically leads to a Yang-Mills-Chern-Simons theory in a curved five-dimensional background space-time whose 
solitons are identified with the baryons of the theory. However, since the soliton solution is at present unknown, one usually argues that when the Chern-Simons coupling and hence 
the size of the of the soliton are sufficiently small, it is possible to invoke the flat space Yang-Mills instanton approximation. The resulting 3+1-dimensional effective theory, obtained via a Kaluza-Klein 
expansion of the holographic direction, contains the Skyrme field and an infinite tower of massive vector mesons. Such holographic theories are a realization of hidden local symmetry where the couplings of the vector mesons are fixed by the Kaluza-Klein mechanism. This extended Skyrme model reduces to the standard Skyrme model upon neglecting all the massive vector meson modes. In the next section we review aspects of quantization of the extended holographic Skyrme model which includes 
hidden local symmetry and vector mesons, since it has a stronger phenomenological appeal.  

\section{SU(2) collective coordinates}

We begin by reviewing the moduli space approximation method to quantize slowly moving solitons. As usual \cite{Adkins:1983ya, Bando:1987br, Hata:2007mb,Hashimoto:2008zw}, the fundamental idea is to approximate the slowly moving soliton by a 
classical soliton solution where the $SU(2)$ moduli are replaced by time-dependent collective coordinates, 
\beq
A_M(t,x^M)=VA^{cl}_M(x^M,X^M(t))V^{-1}-iV\partial_MV^{-1},
\label{su2moduli}
\eeq
where $V=V(t,x^M)$ is an $SU(2)$ element and $X^M$, $M=\{1,\ldots, 4\}$, represents the position of the soliton in the spatial $\mathbb{R}^4$. Introducing the collective coordinates $\ba(t)=a_4(t)+ia_a(t)\tau^a$ as a point in $S^3$ representing the $SU(2)$ orientation, where $a_I=(a_1,a_2,a_3,a_4)$ is a unit vector in $\bR^4$, we note $V(t,x)\rightarrow \ba(t)\; {\rm as}\;x_4\rightarrow \infty$. 
The classical BPST instanton configuration \cite{Hata:2007mb} is given by 
\begin{equation}
A^{cl}_M(x^M)= -i\, \frac{\xi^2 }{\xi^2 + \rho^2} g \del_M g^{-1} \; ,\;g(x^M)=\frac{(x^4-X^4)-i(\vec{x}-\vec X)\cdot\vec{\tau}}{\xi}. 
\label{eq:BPST}
\end{equation}
Let us consider the part of the non-relativistic Hamiltonian related to the collective coordinates obtained from the instanton quantization,
\beq
H=\sum_{I=1}^4 \left(
-\frac{1}{2m}\frac{\del^2}{\del (\rho a_I)^2}+
\half m\omega_\rho^2\rho^2 a_I^2\right)
= \frac{1}{2m}\left(\frac{1}{\rho^3}\partial_\rho(\rho^3\partial_\rho)+\frac{1}{\rho^3}(\nabla^2_{S^3})\right),
\eeq 
where $\rho$ is the radius and $m$ is the mass of the instanton.
The eigenstates (wavefunctions) are factorized into radial and spherical harmonics components. On the $S^3$, they are scalar spherical harmonics and are known to be homogenous polynomials
\beq
T^{(l)}(a_I)=C_{I_1\cdots I_l}\,a_{I_1}\cdots a_{I_l}\ ,
\eeq
where $C_{I_1\cdots I_l}$ is a traceless (in any 2 indexes) symmetric tensor of rank
$l$. They satisfy
\begin{equation}
\nabla^2_{S^3}T^{(l)}=-l(l+2)T^{(l)}\ ,
\end{equation}
The dimension of the tensor $C$ and therefore the dimension of the space of spherical harmonics of degree $l$, $\bH_l$ is $(l+1)^2$. The space $\bH_l$ is a representation of the rotation group $SO(4)$ and corresponds  to the $(S_{l/2},S_{l/2})$ representation of
$(SU(2)\times SU(2))/\bZ_2\simeq SO(4)$.  $S_{l/2}$ denotes the spin $l/2$
representation of $SU(2)$, with $\dim S_{l/2}= l+1$.
Under a group rotation $\ba$ transforms as
\begin{equation}
\ba \to g_I\,\ba\, g_J \ ,\quad
g_{I,J}\in SU(2)_{I,J} \ .
\label{gag}
\end{equation}
where $SU(2)_I$ and $SU(2)_J$ are identified with the
isospin rotation and the spatial rotation, respectively. In order to shed some light on this physical interpretation, we note from equation (\ref{su2moduli})  that $SU(2)_I$ isospin rotation of the gauge fields corresponds to
\begin{equation}
V\to g_I V\ ,\quad g_I\in SU(2)_I\ .
\end{equation}
Under spatial $SO(3)$ rotation, the coordinates change according to
\beq
x^i\rightarrow x'^i= \Lambda^i_jx^j.
\eeq
Hence, 
\beq
g(x)\rightarrow g( \Lambda x)=u^{-1}g(x)u \;,\;\;u\in SU(2),
\label{grotation} 
\eeq
where we used the fact that there is a homomorphism between $SO(3)$ and $SU(2)$ groups and every $SO(3)$ transformation $\Lambda$ in $\bR^3$ corresponds 
to a $SU(2)$ transformation $u$ in the space of anti-hermitian matrices $i\vec x.\vec \tau$ as in  (\ref{grotation}).
Under spatial rotation $\Lambda$ applied in (\ref{eq:BPST}),
\beqa
A^{cl\,i}(x)\rightarrow A'^{cl\,i}(x')&\propto& g( \Lambda x)\Lambda^i_j\partial^jg( \Lambda x)^{-1}=u^{-1}g(x)u\Lambda^i_j\partial^ju^{-1}g(x)^{-1}u\nn \\ 
&=&\Lambda^i_ju^{-1}g(x)\partial^jg(x)^{-1}u. 
\label{clA}
\eeqa
where in the last line we made use of the fact that the $u$ matrix does not depend on $x$ .
With respect to the moduli, the gauge field transforms as
\beqa
&A^i(t,x)&\rightarrow A'^i(t,x')=\Lambda^i_jA^j(t,x)=\Lambda^i_j(VA^{cl\,j}(x)V^{-1}-iV\partial^jV^{-1})\nn\\
&=&V'A'^{cl\,i}(x')V'^{-1}\hspace{-1mm}-iV'\Lambda^i_j\partial^jV'^{-1}=\Lambda^i_j(V'u^{-1}g(x)\partial^jg(x)^{-1}uV'^{-1}\hspace{-1mm}-iV'\partial^jV'^{-1})\nn\\
&=&\Lambda^i_j(V'u^{-1}A^{cl\,j}(x)uV'^{-1}\hspace{-1mm}-iV'\partial^jV'^{-1}).\label{Ai}
\eeqa
where in the first line it should be noted that $A^i$ transforms as a vector and in the second and third lines we used (\ref{grotation}) and (\ref{clA}). Comparing the first and third lines of eq.~(\ref{Ai}), we find that
\beq
V'(t,x')u^{-1}=V(x).
\eeq
Identifying $u=g_J \in SU(2)_J$, we find that $V$ transforms under spatial rotation $SO(3)\simeq SU(2)_J$,
\begin{equation}
V\to V'=Vg_J \ ,\quad g_J\in SU(2)_J\ .
\end{equation} 

\subsection{Non-relativistic baryons}

The instanton collective state is quantized considering slowly moving instantons. Therefore it is related to non-relativistic baryons. We will extend this analysis to relativistic baryons by simply treating them as static instantons boosted in a given direction, disregarding further relativistic corrections. 
First, we relate the spherical harmonics tensors to static nucleons. The lowest states are at $l=1$ and the tensors become linear in $a_I$ coordinates. They correspond to states with spin and isospin $1/2$ and we identify them with protons and neutrons. In spinorial notation we write the particle states as
\beq
|N,h\rangle= \chi^N \otimes\chi_h =: \chi^N_h,
\eeq
where $N=\{p,n\}$, $h=\{+,-\}$ and
\beq
\chi^{p}=\chi_{+} =\left(\begin{array}{c}1\\0\end{array}\right) \quad , \quad \chi^{n}=\chi_{-} =\left(\begin{array}{c}0\\1\end{array}\right).
\label{spinorial notation}
\eeq
The isospin $I_3$ and spin $J_3$ operators in this representation read
\begin{eqnarray}
I_a=\frac{i}{2}\left(
a_4\frac{\del}{\del a_a}-a_a\frac{\del}{\del a_4}
-\epsilon_{abc}\,a_b\frac{\del}{\del a_c}
\right)\ ,~~
J_a=\frac{i}{2}\left(
-a_4\frac{\del}{\del a_a}+a_a\frac{\del}{\del a_4}
-\epsilon_{abc}\,a_b\frac{\del}{\del a_c}
\right),
\label{IJ}
\end{eqnarray}
and their eigenstates are given by
\beqa
|p,+\rangle&=&\frac{a_1 + i a_2}{\pi} \quad, \quad |p,-\rangle=-\frac{i}{\pi}(a_4-i a_3), \nn\\
|n,+\rangle&=&\frac{i(a_4 + i a_3)}{\pi} \quad, \quad |n,-\rangle=-\frac{1}{\pi}(a_1-i a_2). 
\label{nucleonwavefunctions}
\eeqa
One of the objectives of the subsequent section is to generalize these expressions to the relativistic case.

\subsection{Nucleon relativistic wavefunctions}

Initially, we will restrict the discussion to proton and neutron states separately and disregard the isospin information. 
As mentioned before, the relativistic approach presented here does not involve any correction to the non-relativistic effective Lagrangian of the instanton 
collective modes. We consider the static baryons as spin half particles living in Minkowski space and boost them in the $\bp$ direction,
 but there is no explicit knowledge of the other generators of Lorentz symmetry. Regardless of its limitations at high energies, we consider this approach 
to be relevant for calculating the electromagnetic current vacuum expectation values at low energy when baryons have different momenta (inelastic scattering) and 
their helicity states are defined differently under the $\vec p.\vec\sigma$ operator. In such a way, we obtain a description equivalent to
 Dirac spinor states for the baryon wavefunctions that allows us to calculate the matrix elements of baryonic currents and define the corresponding form factors.
We introduce a relativistic spinor for a fermion with a given momentum $\vec p$, 
\beqa
u(p,h) = \frac{1}{\sqrt{2E}} \left( \begin{array}{c} f \chi_h \\ \frac{\vec{p} \cdot \vec{\sigma}}{f} \chi_h \end{array} \right)\quad,\; {\rm with}  \; f = \sqrt{E + m_B} .
\label{u}
\eeqa
Here $\chi_h$ is a helicity eigenstate, which means $\vec{p} \cdot \vec{\sigma}\chi_h=h |\vec p| \chi_h$ where $h= \pm 1$. The action of the operator $\vec{p} \cdot \vec{\sigma}$ on these states 
is given by
\beq
\vec{p} \cdot \vec{\sigma}\,\chi_h = \left(\begin{array}{cc}p_3&p_1-i p_2\\p_1+i p_2&-p_3\end{array}\right)\, \chi_h=h |p| \chi_h.
\label{helicitystates}
\eeq
This yields the eigenstates
\beq
\chi_{+}(\vec p) =\frac{1}{\sqrt{2|\vec p|(|\vec p|+p_3)}}\left(\begin{array}{c}|\vec p|+p_3\\p_1+i p_2\end{array}\right) \quad , \quad \chi_{-}(\vec p) =\frac{1}{\sqrt{2|\vec p|(|\vec p|+p_3)}}\left(\begin{array}{c}-p_1+i p_2\\|\vec p|+p_3\end{array}\right).  
\label{helicity spinors}
\eeq
Notice that  $\chi_h^{\dagger}\chi_h=1$ and $|\vec p|= \sqrt{p_1^2+p_2^2+p_3^3}$. If we choose $\vec{p}=p_3\hat{z}$, we recover $\sigma_3$ eigenstates 
\beq
\chi_{+} =\left(\begin{array}{c}1\\0\end{array}\right) \quad , \quad \chi_{-} =\left(\begin{array}{c}0\\1\end{array}\right).   
\eeq
Hence we expect the proton and neutron linear ($l=1$) spherical harmonic tensors (equivalent to $\chi_h(\vec{p})$ helicity eigenstates) to be
\beqa
\chi_+^{p}(a_i,\vec{p})&=& \frac{(|\vec p|+p_3)(a_1+i a_2)-i(p_1+i p_2)(a_4-i a_3)}{\pi\sqrt{2|\vec p|(|\vec p|+p_3)}},\nn \\
\chi_-^{p}(a_i,\vec{p})&=& \frac{(-p_1+i p_2)(a_1+i a_2)-i(p_3+|\vec p|)(a_4-i a_3)}{\pi\sqrt{2|\vec p|(|\vec p|+p_3)}},\nn\\
\chi_+^{n}(a_i,\vec{p})&=& \frac{(|\vec p|+p_3)i(a_4 + i a_3)-(p_1+i p_2)(a_1-i a_2)}{\pi\sqrt{2|\vec p|(|\vec p|+p_3)}},\nn \\
\chi_-^{n}(a_i,\vec{p})&=& \frac{(-p_1+i p_2)i(a_4 + i a_3)-(p_3+|\vec p|)(a_1-i a_2)}{\pi\sqrt{2|\vec p|(|\vec p|+p_3)}}.
\label{helicity wavefunctions}
\eeqa
The expression above is not valid for $p_1=p_2=0$ and $p_3 = -p$ and in that case we recall that helicity changes sign when momentum changes in sign to
 the opposite direction. This means that
\beq
\chi_h^{N}(a_i,-\vec{p})\equiv \chi_{-h}^{N}(a_i,\vec{p}),
\label{opposite direction}
\eeq
up to a complex phase. As a useful example we write below the case where $\vec p= - |\vec p|\hat{z}$:
\beqa
\frac{\bp}{|\bp|}.\vec J \,\chi_h^{N}(a_i,\vec{p})= h\,\chi_h^{N}(a_i,\vec{p})\nn\\
-\vec J_3 \,\chi_h^{N}(a_i,- |\vec p|\hat{z})=h\,\chi_h^{N}(a_i,-|\vec p|\hat{z})
\eeqa
Therefore $\chi_h^{N}(a_i,- |\vec p|\hat{z})$ is an eigenstate with eigenvalue $-h$ of the $+\hat{z}$ direction operator $J_3$:
\beq
\chi_h^{N}(a_i,- \hat{z}) = \chi_{-h}^{N}(a_i,+\hat{z})
\eeq 

After some algebra we can verify that (\ref{helicity wavefunctions}) are eigenfunctions of $\bp. \vec J$ operators (\ref{IJ}):
\beq
\bp.\vec J\,\chi_+^{N}(a_i,\vec{p})=+\frac{1}{2}|\vec p| \,\chi_+^{N}(a_i,\vec{p})\quad {\rm and} \quad \vec p.\vec J\,\chi_-^{N}(a_i,\vec{p})=-\frac{1}{2}|\vec p| \,\chi_-^{N}(a_i,\vec{p}).
\eeq
In the static case the nucleon wavefunctions (\ref{nucleonwavefunctions}), (\ref{helicity wavefunctions}) represent 2-spinor helicity states, cf. (\ref{spinorial notation}). Therefore the complete relativistic nucleon SU(2) wavefunctions are 2-spinors containing as components the helicity states $\chi_{\pm}^{N}(a_i,\vec p)$ (SU(2) wavefunctions); thus these spinors are equivalent to Dirac 4-spinors, cf. (\ref{u}). The nucleon SU(2) wavefunction becomes
\beq
u_N(a_i,\bp,+)= \frac{1}{\sqrt{2E}} \left( \begin{array}{c} f \chi_+^{N}(a_i,\vec p) \\ \frac{|\vec{p}|}{f} \chi_+^{N}(a_i,\vec p) \end{array} \right) \quad , \quad 
u_N(a_i,\bp,-) = \frac{1}{\sqrt{2E}} \left( \begin{array}{c} f \chi_-^{N}(a_i,\vec p) \\ -\frac{|\vec{p}|}{f} \chi_-^{N}(a_i,\vec p)  \end{array} \right) \, .
\eeq
The explicit dependence on $a_I \in S^3$ is not directly observable but rather encodes the spin/isospin in this representation. The integration in $S^{3}$ moduli recovers the 4D Dirac spinor with appropriate isospin, as we will show in the next subsection.

\subsection{Dirac equation and spin sum}
 
In order to work with relativistic $SU(2)$ wavefunctions instead of Dirac spinorial notation, we define the substitutes of gamma matrices in the $SU(2)$ collective space by simply replacing the Pauli matrices $\sigma^i$ with $2 J^i$ operators (\ref{IJ}).  Hence, the new $2\times2$ gamma matrices are
\beq
\gamma^0=-i\left( \begin{array}{cc} 1&0 \\ 0&-1 \end{array} \right) \quad , \quad 
\gamma^i=-i\left( \begin{array}{cc} 0&2J^i \\ -2J^i &0 \end{array} \right)  \, .
\eeq
Such operators act only on spin and we will disregard the isospin index for now. Upon such substitution we can verify the validity of the Dirac equation:
\beq
(i  \!\psl+m_B)u_N(a_i,\bp,h)=0.
\eeq
Since $J^i$ operators have real eigenvalues and behave like Pauli matrices, we define their operation to the left by transpose conjugation:
\beq
\psi_h^\dagger (\vec p. \vec J)=(\vec p . \vec J \psi_h)^\dagger = |\vec p| \frac{h}{2} \psi^\dagger_h
\eeq
and using $\ol u_N(a_i,\bp, h)=u_N^\dagger(a_i,\bp, h)i\gamma^0 $ we get the second Dirac equation:
\beq
\ol u_N(a_i,\bp, h) (i  \!\psl+m_B)= 0.
\eeq
As mentioned before, the spinor normalization is given by an integration of the $a_i$ moduli,
\beq
\ol u(\bp, h')u(\bp, h)= \int_{S^3}\ol u(a_i,\bp, h')u(a_i,\bp, h)
\eeq
Working out the integrand,
\beqa
 \ol u(a_i,\bp, h')u(a_i,\bp, h)&=& \frac{1}{2E}(f\chi_{h'}^{*}(a_i,\bp)\quad \tfrac{|\bp|}{f}h'\chi_{h'}^*(a_i,\bp))\left( \begin{array}{cc} 1&0 \\ 0&-1 \end{array} \right) \left( \begin{array}{c} f \chi_h^{a_i}(\vec p) \\ \frac{|\vec{p}|}{f}h \chi_h(a_i,\vec p) \end{array} \right) \nn\\
&=& \frac{1}{2E}\left(f^2-h'h\frac{|\bp|^2}{f^2}\right)\chi_{h'}^{*}(a_i,\bp)\chi_h(a_i,\bp)
\label{oluuintegrand}
\eeqa
In order to integrate (\ref{oluuintegrand}) over $S^3$ we write $(a_1,a_2,a_3,a_4)$ in spherical coordinates:
\beqa
a_1&=&\sin(\theta_0)\sin(\theta_1)\sin(\theta_2),\nn\\
a_2&=&\sin(\theta_0)\sin(\theta_1)\cos(\theta_2),\nn\\
a_3&=&\sin(\theta_0)\cos(\theta_1),\nn\\
a_4&=&\cos(\theta_0).
\eeqa
The volume element is $d\Omega_3 = \sin^2\theta_0 \sin\theta_1
d\theta_0 d\theta_1 d\theta_2$. Using (\ref{helicity wavefunctions}), the integration over (\ref{oluuintegrand}) turns out to be
\beq
\int_{S^3} \ol u(a_i,\bp, h')u(a_i,\bp, h) =\frac{1}{2E} \left((E+m_B)-h'h(E-m_B)\right)\int_{S^3}\chi_{h'}^{a_i*}(\bp)\chi_h^{a_i}(\bp)=\frac{m_B}{E}\delta_{h'h}.
\label{unorm}
\eeq
The spin sum is given by
\beq
\sum_hu(\bp,h)\ol u(\bp,h)=\sum_h\int_{S^3}u(a_i,\bp,h)\ol u(a_i,\bp,h), 
\label{spinsum}
\eeq
where the integrand of (\ref{spinsum}) reads
\beqa
u(a_i,\bp,h)\ol u(a_i,\bp,h) &=&  \frac{1}{2E}\left( \begin{array}{c} f \chi_h^{a_i}(\vec p) \\ \frac{2\vec J.\bp}{f}\chi_h^{a_i}(\vec p) \end{array} \right)(f\chi_h^{a_i*}(\bp)\quad -\tfrac{2\vec J.\bp}{f}\chi_h^{a_i*}(\bp)) \nn\\
&=& \frac{1}{2E}\left(\begin{array}{cc}E+m_B&-2\vec J.\bp\\2\vec J.\bp&-E+m_B\end{array}\right)|\chi_h^{a_i}(\bp)|^2 \nn \\ &=& \frac{1}{2E}(-i  \!\psl+m_B)|\chi_h^{a_i}(\bp)|^2.
\label{spinsumintegrand}
\eeqa
Integrating over (\ref{spinsumintegrand}) we get the spin sum (\ref{spinsum}),
\beqa
\sum_h\int_{S^3}u(a_i,\bp,h)\ol u(a_i,\bp,h)=\frac{1}{2E}(-i  \!\psl+m_B)\sum_h\int_{S^3}|\chi_h^{a_i}(\bp)|^2=\frac{1}{E}(-i  \!\psl+m_B).
\eeqa

\section{Application: vacuum expectation values of traces and elastic form factors}

In \cite{Adkins:1983ya}, the authors calculate the proton isovector magnetic moment which originates from the vector current $J_V=J_L+J_R$ of $SU(2)_L \times SU(2)_R$. 
An important ingredient in the derivation is the calculation of the vacuum expectation values (vevs) of  $\Tr(\tau^i\ba^{-1}\tau^3\ba)$ for static protons. 
This is related to the fact that the electromagnetic current is a linear combination of the isoscalar and isovector currents, i.e., $\mathcal{J}^{\mu}:= \frac{1}{N_c} J^{\mu,0}_V +  J^{\mu,3}_V$.
Here, we extend this calculation to external moving protons, and apply it subsequently in the derivation of elastic form factors in a general framework with vector meson dominance.

\subsection{Trace vevs}
In spherical coordinates the traces become
\beqa
\Tr(\tau^1\ba^{-1}\tau^3\ba)&=&4 \sin(\theta_0) \sin(\theta_1) (\cos(\theta_0) \cos(\theta_2)+ \cos(\theta_1) \sin(\theta_0) \sin(\theta_2)),\nn\\
\Tr(\tau^2\ba^{-1}\tau^3\ba)&=&4 \sin(\theta_0) \sin(\theta_1) (\cos(\theta_1) \cos(\theta_2)\sin(\theta_0)- \cos(\theta_1)  \sin(\theta_2)),\nn\\
\Tr(\tau^3\ba^{-1}\tau^3\ba)&=&2 (\cos(\theta_0)^2 + \cos(2 \theta_1) \sin(\theta_0)^2).
\eeqa
Rewriting the proton states in this notation, we get
\beqa
\langle  \Tr(\tau^a\ba^{-1}\tau^3\ba)\rangle&\equiv&\langle B_X, \bp', h' |\Tr(\tau^a\ba^{-1}\tau^3\ba)|B_0,\bp,h \rangle\nn\\
&=& \int_{S^3}d\Omega^3\;\ol u_X(a_i,\bp',h')\Tr(\tau^a\ba^{-1}\tau^3\ba)u_0(a_i,\bp,h)
\eeqa
The $S^3$ integration above simplifies if we write the helicity states $\chi_{h'}(a_i,\bp')$  and $\chi_h(a_i,\bp)$ as linear combination of the basis  $\{\chi_+(a_i),\chi_-(a_i) \}$ of helicity states of $z$ direction as in eq. (\ref{nucleonwavefunctions}). The integration is resummed to a linear combination of the following form (as in \cite{Adkins:1983ya}):
\beq
\int_{S^3}d\Omega^3\chi^*_{h'}(a_i)\Tr(\tau^a\ba^{-1}\tau^3\ba)\chi_h(a_i)=- \frac{2}{3}\sigma^a_{h'h}=- \frac{2}{3} \chi^\dagger_{h'}\sigma^a\chi_h
\label{staticvev}
\eeq
The helicity states can be written as:
\beq
\chi_h(a_i,\bp)=\frac{(|\bp|+p_3)\chi_h(a_i)+(hp_1+ip_2)\chi_{-h}(a_i)}{\sqrt{2|\vec p|(|\vec p|+p_3)}}
\eeq
%
Thus we find,
\beq
\int_{S^3}d\Omega^3\;\ol u_X(a_i,\bp',h')\Tr(\tau^a\ba^{-1}\tau^3\ba)u_0(a_i,\bp,h)=\frac{-1}{3\sqrt{E_XE}}(ff_X-\tfrac{hh'|\bp||\bp'|}{ff_X})\chi^\dagger_{h'}(\bp')\sigma^a\chi_h(\bp)
\label{general trace}
\eeq
where $f_X=\sqrt{E_X+m_{B_X}}$ and $\chi_{h'}(\bp')$ and $\chi_h(\bp)$ are the helicity spinors (\ref{helicity spinors}). 
The equation above plays an important role in the calculation of the electromagnetic proton form factors \cite{Sakai:2004cn}. In the elastic case, and in the Breit frame
with $\bp=-\bp'=-\dfrac{q}{2}\hat z$, the expression (\ref{general trace}) takes the form
\beqa
\langle \Tr(\tau^a\ba^{-1}\tau^3\ba)\rangle &=& \frac{1}{2E}\left(E+m_B-h'h(E-m_B)\right)\left(-\frac{2}{3}\right)\sigma^a_{h', -h} \nonumber \\&=&\left\{ \;\begin{array}{c} \; \left(-\frac{2}{3}\right) \frac{m_B}{E}\sigma^1_{h', -h}\; {\rm for }\; a=1.\\ \;\left(-\frac{2}{3}\right) \frac{m_B}{E}\sigma^2_{h', -h}\; {\rm for }\; a=2.\\ \;\left(-\frac{2}{3}\right)\sigma^3_{h', -h}\; {\rm for }\; a=3.  \end{array}\right.
\label{zcorrection}
\eeqa
where we used the substitution (\ref{opposite direction}). This amounts to a relativistic correction in the direction of movement $\hat z$ w.r.t.~the static case \cite{Adkins:1983ya}, cf. (\ref{staticvev}), of a factor of $\frac{m_B}{E}$. 

\subsection{SU(2) vector current matrix elements and elastic form factors}

In this section we will demonstrate in a general example how eq.~(\ref{general trace}) can provide a relativistic correction to SU(2) current vevs. 
Considering first the pure Skyrme models, the authors of \cite{Adkins:1983ya} (cf.~section 2), mention that a rather involved calculation has to be carried out in order to find the SU(2) vector current in the Skyrme model. 
In our approach, however, an integration of the currents over the unit sphere of spatial $\mathbb R^3$ shows that
\beq
\int d \Omega^2 J^{0,c}_V\sim \; \Tr[(\partial_0\ba)\ba^{-1}\tau^c]\, , \quad {\rm and} 
\quad \eps^{ijk}\hspace{-2mm}\int d \Omega^2 x^j J^{k,c}_V\sim \; \Tr[\tau^i \ba^{-1}\tau^c\ba],
\eeq
where it should be noted that the $SU(2)_V$ currents were decomposed as $J^\mu_V=J^{\mu,c}_V\tau^c$.
Since the collective coordinates do not depend on the position in $\mathbb R^3$, general vector currents read
\beqa
  J_V^{0,c}(\bk)&=&e^{-i\bk\cdot\bX}I_c
\label{wtJ0}\\
  J_V^{i,c}(\bk)&=&ie^{-i\bk\cdot\bX}\,\Lambda
\epsilon_{ija}q_j\Tr\left(\tau^c\ba\tau^a\ba^{-1}\right)
\label{wtJj}
\eeqa
where $q_j := p'_j - p_j$. The constant\footnote{The constant $\Lambda$ is model-dependent and will not be fixed to a specific value here. 
In order to compare our results with the Sakai-Sugimoto model, one should set $\Lambda = 2 \pi^2 \kappa \langle \rho^2 \rangle$.} $\Lambda$ has dimension $[{\rm mass}]^{-1}$, and 
in (\ref{wtJj}) we have used the canonical momentum defined in \cite{Adkins:1983ya}, i.e.,
\beq
\partial_0 a_i= \frac{\pi_i}{4\lambda}=\frac{-i}{4\lambda}\frac{\partial}{\partial a_i}. 
\eeq
Therefore we obtain
\beq \Tr[(\partial_0\ba)\ba^{-1}\tau^c]=\frac{I_c}{2\lambda},
\eeq
according to (\ref{IJ}), where we have introduced $\lambda := \frac{2\pi }{3 e^3 f_{\pi}}\Lambda$, cf.~\cite{Adkins:1983ya}, in terms of the model-dependent pion decay constant $f_{\pi}$ and 
dimensionless Skyrme parameter $e$.\\
When calculating the vector current proton-proton matrix elements, we are only interested in the $c=3$ components (see above). 
We define the Dirac and Pauli form factors according to the following decomposition of current matrix elements, cf.~\cite{Hashimoto:2008zw,Bayona:2011xj},
\beqa \label{currentdecomp}
\langle p_{\ss  X} , B_{\ss  X} , s_{\ss  X}  \vert J_V^{\mu,a} (0) \vert p , B, s \rangle &=& \frac{i}{2 (2 \pi)^3}  (\tau^a)_{I_3^{\ss  X} I_3}
\left ( \eta^{\mu \nu}  - \frac{ q^\mu q^\nu}{q^2} \right )
 \bar u (p_{\ss  X}  , s_{\ss  X}) \Big [  \gamma_\nu F^{D,a}_{B B_{\ss  X}}(q^2)  \cr
&& \quad + \kappa_B  \sigma_{\nu \lambda} q^\lambda  F^{P,a}_{B B_{\ss  X}} (q^2)  \Big ] u (p , s)  \, .
\eeqa
Note that this expression contains a projector to ensure transversality of the currents $\partial_\mu\langle J^\mu\rangle=0$, as the currents are defined from a non-relativistic theory. 
In order to compare our results with the literature, we find it convenient to restrict to the elastic case in the Breit frame, with $\bp=-\bp'=-\dfrac{q}{2} \hat z$. 
In this case, the expression (\ref{currentdecomp}) becomes 
\beqa
 \hspace{-3mm}\langle p_{\ss  X} , B_{\ss  X} , s_{\ss  X}  \vert J_V^{0,a} (0) \vert p , B, s \rangle
&=& \frac{1}{2 (2 \pi)^3}  (\tau^a)_{I_3^{\ss  X} I_3} \left ( \frac{m_B}{E} \right ) \delta_{s_{\ss X},- s}
\left [ F^{D,a}_{B}(q^2) - \frac{q^2}{4 m_B^2} F^{P,a}_{B}(q^2) \right ], \cr
&\equiv& \frac{1}{2 (2 \pi)^3}  (\tau^a)_{I_3^{\ss  X} I_3} \left ( \frac{m_B}{E} \right ) \delta_{s_{\ss X},- s}
G^{E,a}_{B}(q^2) \, ,
\label{sachse}
\eeqa
\beqa
\hspace{-10mm}\langle p_{\ss  X} , B_{\ss  X} , s_{\ss  X}  \vert J_V^{i,a} (0) \vert p , B, s \rangle
&=&  -\frac{1}{2 (2 \pi)^3}  (\tau^a)_{I_3^{\ss  X} I_3} \left ( \frac{i}{2E} \right )
\epsilon^{ijk} q_j (\sigma_k)_{s_{\ss  X},- s}
\left [ F^{D,a}_{B}(q^2) + F^{P,a}_{B}(q^2) \right ], \cr
&\equiv &- \frac{1}{2 (2 \pi)^3}  (\tau^a)_{I_3^{\ss  X} I_3} \left ( \frac{i}{2E} \right)
\epsilon^{ijk} q_j (\sigma_k)_{s_{\ss  X},- s}
G^{M,a}_{B}(q^2) \, .
\label{sachsm}
\eeqa
The elastic form factors $G^{E,a}_{B}(q^2)$ and $G^{M,a}_{B}(q^2)$ are known as the {\it Sachs form factors}.\\ 
The aforementioned expressions include a relativistic correction that did not naturally occur in many other generalizations of Skyrme (-type) models (cf.~\cite{Panico:2008it}, chapter 3.2). 
This is the main advantage of our prescription.\\
Using the vector currents defined in (\ref{wtJ0}) and (\ref{wtJj}) and the result (\ref{zcorrection}), the current matrix elements can be written as
 \beqa
\langle \dfrac{q}{2} \hat z, B_{\ss  X} , h'  \vert J_V^{0,3} (0) \vert-\dfrac{q}{2} \hat z, B, h \rangle &=& 
\frac{1}{2 (2 \pi)^3} \left(\frac{m_B}{E}\right) \delta_{-h'h}  \label{Je0}\\
\langle \dfrac{q}{2} \hat z , B_{\ss  X} , h'  \vert J_V^{i,3} (0) \vert -\dfrac{q}{2} \hat z , B, h \rangle &=& -\frac{1}{2(2 \pi)^3} \left(\frac{i}{2E}\right) \frac{8m_B}{3} \Lambda 
\epsilon_{i3a} q_3 \sigma^a_{-h'h}
\label{Jej}
\eeqa
where we utilized eq.~(\ref{unorm}) and noted that $\langle 2 I^3 \rangle_{+1/2,+1/2}=(\tau^3)_{+1/2,+1/2}=1$ for the proton-proton matrix elements.\\
Employing eqs.~(\ref{Je0}), (\ref{Jej}) and eqs.~(\ref{sachse}), (\ref{sachsm}) we arrive at
\beqa
 F_B^{D,3}(q^2)&=&\left( 1 + \frac{q^2}{4 m_B^2} \right)^{-1} \left( \frac{2}{3}\frac{\Lambda q^2}{m_B}  + 1\right),\\
 F_B^{P,3}(q^2)&=&\left( 1 + \frac{q^2}{4 m_B^2} \right)^{-1} \left( \frac{8}{3}m_B \Lambda -1 \right).
\eeqa
In models which exhibit vector meson dominance, the diagonal $SU(2)$ symmetry of the chiral symmetry $SU(2)_L \times SU(2)_R$ is gauged and thus the diagonal $SU(2)$ current couples to the gauge field. 
In a na{\"i}ve approach, the complete currents can be constructed by including appropriate factors from the $\rho$-meson couplings to the gauge field ($g_{\rho}$), protons ($g_{\rho BB}$) and its propagator. \\
Accordingly, for the case of $\rho$-meson dominance, the result for the Dirac and Pauli form factors reads
 \beqa
F^{D,3}_{\rho \rm{MD}}(q^2)&=& \left( 1 + \frac{q^2}{4 m_B^2} \right)^{-1} \left( \frac{2}{3}\frac{\Lambda q^2}{m_B}  + 1\right)\;\frac{g_{\rho}g_{\rho BB}}{m_\rho^2+q^2},\\
F^{P,3}_{\rho \rm{MD}}(q^2)&=&\left( 1 + \frac{q^2}{4 m_B^2} \right)^{-1} \left( \frac{8}{3}m_B \Lambda -1 \right)\; \frac{g_{\rho}g_{\rho BB}}{m_\rho^2+q^2}.
\eeqa 
It should be noted that our results for the Dirac and Pauli form factor derived above  (or equivalently for the Sachs form factors), are in agreement with the form factors given in \cite{Hashimoto:2008zw} when setting $\Lambda := 2 \pi^2 \kappa \langle \rho^2 \rangle$.
\section{Conclusion}
In this note, we have presented a relativistic generalization for non-static baryons in certain Skyrme-type models. This allows us to overcome an old problem concerning the correct normalization of skyrmions: In \cite{Panico:2008it}, the authors mention this limitation and adopt the spinor 
normalization of the static case in the calculation of the current matrix elements (see also \cite{Park:2008sp, Grigoryan:2009pp}). In \cite{Hashimoto:2008zw}, the authors {\it impose} the relativistic spinor normalization before calculating 
the current matrix elements. The correct relativistic normalization for currents and the vacuum expectation values appears 
quite naturally in our approach to skyrmions. 
A more involved application of our approach is the case of non-elastic scattering, where relativistic corrections redefine the current matrix elements and form factors. This was studied in detail in \cite{Bayona:2011xj}.  

\section*{Acknowledgements}
The authors are indebted to C.A.~Ballon Bayona for collaboration in the early stages of the project and would like acknowledge useful correspondence with S.~Sugimoto.
The authors H.B-F., N.R.F.B. and M.A.C.T. are partially supported by CAPES and CNPq (Brazilian research agencies). The work of M.I. was supported by an IRCSET postdoctoral fellowship.


\begin{thebibliography}{19}

\bibitem{Skyrme:1961vq}
  T.~H.~R.~Skyrme,
  ``A Nonlinear field theory,''
  Proc.\ Roy.\ Soc.\ Lond.\  A {\bf 260}, 127 (1961).
  
\bibitem{Adkins:1983ya}
  G.~S.~Adkins, C.~R.~Nappi, E.~Witten,
  ``Static Properties of Nucleons in the Skyrme Model,''
  Nucl.\ Phys.\  {\bf B228}, 552 (1983).
\bibitem{Adkins:1983hy}
  G.~S.~Adkins, C.~R.~Nappi,
  ``The Skyrme Model with Pion Masses,''
  Nucl.\ Phys.\  {\bf B233}, 109 (1984).
  
\bibitem{Braaten:1986iw}
  E.~Braaten, S.~-M.~Tse, C.~Willcox,
  ``Electromagnetic Form-factors In The Skyrme Model,''
  Phys.\ Rev.\ Lett.\  {\bf 56}, 2008 (1986).
\bibitem{Holzwarth:2005re}
  G.~Holzwarth,
  ``Electromagnetic form-factors of the nucleon in chiral soliton models,''
  [hep-ph/0511194].

 \bibitem{Sakai:2004cn}
  T.~Sakai and S.~Sugimoto,
  ``Low energy hadron physics in holographic QCD,''
  Prog.\ Theor.\ Phys.\  {\bf 113}, 843 (2005)
  [arXiv:hep-th/0412141].
\bibitem{Hata:2010vj}
  H.~Hata, T.~Kikuchi,
  ``Relativistic Collective Coordinate Quantization of Solitons: Spinning Skyrmion,''
  Phys.\ Rev.\  {\bf D82}, 025017 (2010).
  [arXiv:1002.2464 [hep-th]].
\bibitem{Hata:2010zy}
  H.~Hata, T.~Kikuchi,
  ``Relativistic Collective Coordinate System of Solitons and Spinning Skyrmion,''
  Prog.\ Theor.\ Phys.\  {\bf 125}, 59-101 (2011).
  [arXiv:1008.3605 [hep-th]].
\bibitem{Ji:1991ff}
  X.~-D.~Ji,
  ``A Relativistic skyrmion and its form-factors,''
  Phys.\ Lett.\  {\bf B254}, 456-461 (1991).

\bibitem{Klebanov:1985qi} 
  I.~R.~Klebanov,
  Nucl.\ Phys.\ B {\bf 262}, 133 (1985).

\bibitem{Kugler:1988mu} 
  M.~Kugler and S.~Shtrikman,
  Phys.\ Lett.\ B {\bf 208}, 491 (1988).

\bibitem{Atiyah:1989dq}
 M.~F.~Atiyah, N.~S.~Manton,
 ``Skyrmions From Instantons,''
 Phys.\ Lett.\  {\bf B222}, 438-442 (1989).
\bibitem{Sutcliffe:2010et}
  P.~Sutcliffe,
  ``Skyrmions, instantons and holography,''
  JHEP {\bf 1008}, 019 (2010).
  [arXiv:1003.0023 [hep-th]].
\bibitem{Hata:2007mb}
  H.~Hata, T.~Sakai, S.~Sugimoto, S.~Yamato,
  ``Baryons from instantons in holographic QCD,''
  Prog.\ Theor.\ Phys.\  {\bf 117}, 1157 (2007).
  [hep-th/0701280].

\bibitem{Hashimoto:2008zw}
  K.~Hashimoto, T.~Sakai and S.~Sugimoto,
  ``Holographic Baryons: Static Properties and Form Factors from Gauge/String
  Duality,''
  Prog.\ Theor.\ Phys.\  {\bf 120} (2008) 1093
  [arXiv:0806.3122 [hep-th]].

\bibitem{BallonBayona:2009ar}
  C.~A.~Ballon Bayona, H.~Boschi-Filho, N.~R.~F.~Braga, M.~A.~C.~Torres,
  ``Form factors of vector and axial-vector mesons in holographic D4-D8 model,''
  JHEP {\bf 1001}, 052 (2010).
  [arXiv:0911.0023 [hep-th]].

\bibitem{Bayona:2010bg}
  C.~A.~B.~Bayona, H.~Boschi-Filho, M.~Ihl, M.~A.~C.~Torres,
  ``Pion and Vector Meson Form Factors in the Kuperstein-Sonnenschein holographic model,''
  JHEP {\bf 1008}, 122 (2010).
  [arXiv:1006.2363 [hep-th]].

\bibitem{Ihl:2010zg}
  M.~Ihl, M.~A.~C.~Torres, H.~Boschi-Filho, C.~A.~B.~Bayona,
  ``Scalar and vector mesons of flavor chiral symmetry breaking in the Klebanov-Strassler background,''
  JHEP {\bf 1109}, 026 (2011).
  [arXiv:1010.0993 [hep-th]].

\bibitem{Pomarol:2007kr}
  A.~Pomarol and A.~Wulzer,
  ``Stable skyrmions from extra dimensions,''
  JHEP {\bf 0803} (2008) 051
  [arXiv:0712.3276 [hep-th]].
  
\bibitem{Bando:1987br}
  M.~Bando, T.~Kugo and K.~Yamawaki,
  ``Nonlinear Realization and Hidden Local Symmetries,''
  Phys.\ Rept.\  {\bf 164} (1988) 217.

\bibitem{Finkelstein:1968hy}
  D.~Finkelstein and J.~Rubinstein,
  ``Connection between spin, statistics, and kinks,''
  J.\ Math.\ Phys.\  {\bf 9} (1968) 1762.
  
\bibitem{Panico:2008it}
  G.~Panico and A.~Wulzer,
  ``Nucleon Form Factors from 5D Skyrmions,''
  Nucl.\ Phys.\  A {\bf 825} (2009) 91
  [arXiv:0811.2211 [hep-ph]].
\bibitem{Park:2008sp} 
  J.~Park and P.~Yi,
  ``A Holographic QCD and Excited Baryons from String Theory,''
  JHEP {\bf 0806}, 011 (2008)
  [arXiv:0804.2926 [hep-th]].

\bibitem{Grigoryan:2009pp} 
  H.~R.~Grigoryan, T.~-S.~H.~Lee and H.~-U.~Yee,
  ``Electromagnetic Nucleon-to-Delta Transition in Holographic QCD,''
  Phys.\ Rev.\ D {\bf 80}, 055006 (2009)
  [arXiv:0904.3710 [hep-ph]].
  
\bibitem{Bayona:2011xj}
  C.~A.~B.~Bayona, H.~Boschi-Filho, N.~R.~F.~Braga, M.~Ihl and M.~A.~C.~Torres,
  ``Generalized baryon form factors and proton structure functions in the
  Sakai-Sugimoto model,''
  arXiv:1112.1439 [hep-ph].
\end{thebibliography}
\end{document}